\documentclass[12pt,english]{article}
\usepackage{mathpazo}
\usepackage[latin9]{inputenc}
\usepackage[letterpaper]{geometry}
\usepackage{babel}
\usepackage{array}
\usepackage{mathtools}
\usepackage{bm}
\usepackage{multirow}
\usepackage{amsmath}
\usepackage{amssymb}
\usepackage{graphicx}
\usepackage{setspace}
\usepackage[authoryear]{natbib}
\usepackage[unicode=true]
 {hyperref}

\makeatletter

\providecommand{\tabularnewline}{\\}

\usepackage{amsfonts}
\usepackage{amsthm}
\usepackage{fancyhdr}
\usepackage{lastpage}
\usepackage{extramarks}
\usepackage{chngpage}
\usepackage{soul}
\usepackage{color}
\usepackage{graphicx}
\usepackage{graphics}
\usepackage{lscape}

\usepackage{bm}
\usepackage{enumerate}
\usepackage{verbatim}
\usepackage{babel}
\usepackage{titlesec}

\def\ci{\perp\!\!\!\perp}

\usepackage{pifont}

\titleformat{\subsection}
  {\normalfont\large\bfseries}{\thesubsection}{1em}{}

\usepackage{pgf}
\usepackage{tikz}\usetikzlibrary{positioning}
\usetikzlibrary{arrows}

\usepackage{babel}

\makeatother

\begin{document}

\title{{\Large{}\vspace{-30pt}
}Residual Balancing: A Method of Constructing Weights for Marginal
Structural Models\thanks{Direct all correspondence to Xiang Zhou, Department of Government,
Harvard University, 1737 Cambridge Street, Cambridge, MA 02138, USA;
email: xiang\_zhou@fas.harvard.edu. The authors benefited from communications
with Justin Esarey, Kosuke Imai, Gary King, José Zubizarreta, and
participants of the Applied Statistics Workshop at Harvard University,
the Political Methodology Speaker Series at MIT, and the 35th Annual
Meeting of the Society for Political Methodology at Brigham Young
University. }}

\author{Xiang Zhou\\
Harvard University\and Geoffrey T. Wodtke\\
University of Toronto}

\date{March 29, 2019\vspace{-10pt}}
\maketitle
\begin{abstract}
\noindent When making causal inferences, post-treatment confounders
complicate analyses of time-varying treatment effects. Conditioning
on these variables naively to estimate marginal effects may inappropriately
block causal pathways and may induce spurious associations between
treatment and the outcome, leading to bias. To avoid such bias, researchers
often use marginal structural models (MSMs) with inverse probability
weighting (IPW). However, IPW requires models for the conditional
distributions of treatment and is highly sensitive to their misspecification.
Moreover, IPW is relatively inefficient, susceptible to finite-sample
bias, and difficult to use with continuous treatments. We introduce
an alternative method of constructing weights for MSMs, which we call
\textquotedblleft residual balancing.\textquotedblright{} In contrast
to IPW, it requires modeling the conditional means of the post-treatment
confounders rather than the conditional distributions of treatment,
and it is therefore easier to use with continuous exposures. Numeric
simulations suggest that residual balancing is both more efficient
and more robust to model misspecification than IPW and its variants.
We illustrate the method by estimating (a) the cumulative effect of
negative advertising on election outcomes and (b) the controlled direct
effect of shared democracy on public support for war. Open source
software is available for implementing the proposed method.

\noindent \newpage{}
\end{abstract}
\spacing{1.5}

\section{Introduction}

Social scientists are often interested in estimating the marginal,
or population average, effects of treatment in the presence of post-treatment
confounding. Post-treatment confounding is common in studies of time-varying
treatments, where confounders of future treatments may be affected
by prior treatments. For example, political scientists study how the
timing and frequency of negative advertising during political campaigns
affect election outcomes (e.g., \citealt{lau2007effects,blackwell2013framework}).
In this context, the decision to run negative advertisements at any
given point during a campaign is affected by a candidate's position
in recent polling data, which itself is affected by negative advertising
conducted previously. Post-treatment confounding is also common in
analyses of causal mediation, where confounders for the effect of
the mediator on the outcome may be affected by treatment. For example,
when assessing the role of morality in mediating the effects of shared
democracy on public support for war, post-treatment variables, such
as beliefs about the threat posed by the adversary, may affect both
the perceived morality of war and support for military action (\citealt{tomz2013dempeace}).

Conventional methods that adjust for post-treatment confounders by
conditioning, stratifying, or matching on them naively may engender
two different types of bias (\citealt{robins1986new,robins1999marginal}).
First, adjusting naively for post-treatment confounders leads to bias
from over-control of intermediate pathways because it blocks, or ``controls
away,'' the effect of treatment on the outcome that operates through
these variables. Second, adjusting naively for post-treatment confounders
can lead to collider-stratification bias if these variables are also
affected by unobserved determinants of the outcome, as conditioning
on a variable generates a spurious association between its common
causes even when these common causes are unconditionally independent
(\citealt{pearl2009causality}).

To avoid these biases, researchers typically use marginal structural
models (MSMs) and the associated method of inverse probability weighting
(IPW), which yields consistent estimators of treatment effects under
fairly general conditions (\citealt{robins1999marginal,robins2000marginal,vanderweele2015explanation}).
Nevertheless, IPW is not without limitations. First, IPW requires
models for the conditional distributions of exposure to treatment
and/or the mediator, and prior research indicates that it is highly
sensitive to their misspecification (\citealt{mortimer2005application,kang2007demystifying,lefebvre2008impact,howe2011limitation}).
Second, even if these models are correctly specified, IPW is relatively
inefficient, and it is susceptible to large finite-sample biases when
confounders strongly predict the exposures of interest (\citealt{wang2006diagnosing,cole2008constructing}).\footnote{For expositional simplicity, we occasionally use the term ``exposures''
to generally refer to treatments or mediators. } Finally, when the exposures of interest are continuous, IPW tends
to perform poorly because estimates of conditional densities are often
unreliable (e.g., \citealt{vansteelandt2009estimating,naimi2014constructing}).

Several remedies have been proposed to improve the efficiency and
robustness of IPW. For example, \citet{cole2008constructing} suggest
truncating or censoring extreme weights to obtain more precise estimates.
With this method, however, the improved precision comes at the cost
of greater bias. Recently, \citet{imai2014covariate,imai2015robust}
propose constructing weights for an MSM with covariate balancing propensity
scores (CBPS). By integrating a large set of balancing conditions
when estimating propensity scores, this method is less sensitive to
model misspecification. But estimating CBPS can be computationally
demanding, and because of the practical difficulties associated with
modeling conditional densities, this method is not well suited for
continuous exposures (see \citealt{fong2018covariate} and \citealt{yiu2018covariate}
for extensions of CBPS to continuous exposures in the cross-sectional
setting).

In this paper, we propose an alternative method of constructing weights
for MSMs, which we call ``residual balancing.'' Briefly, the method
is implemented in two stages. First, a model for the conditional mean
of each post-treatment confounder, given past treatments and confounders,
is estimated and then used to construct residual terms. Second, a
set of weights is constructed using \citeauthor{hainmueller2012entropy}'s
(2012) entropy balancing method such that, in the weighted sample,
(a) the residualized confounders are orthogonal to future exposures,
past treatments, and past confounders, and (b) their discrepancy with
a set of base weights (e.g., survey sampling weights) is minimized.
Thus, our proposed method is an extension of \citeauthor{hainmueller2012entropy}'s
(2012) entropy balancing procedure to the longitudinal setting. It
exactly balances sample moments for each of the post-treatment confounders
across future exposures, conditional on the observed past, without
explicit models for the conditional distributions of exposure to treatment
and/or a mediator. 

This method has a number of advantages over IPW and its variants.
First, residual balancing is relatively robust to the model misspecification
bias that commonly afflicts these other methods. Second, residual
balancing is also more efficient because it tends to avoid highly
variable and extreme weights by minimizing their relative entropy
with respect to a set of base weights. Third, because it does not
require models for the conditional distributions of the exposures,
residual balancing is easy to use with continuous treatments and/or
mediators. Finally, in contrast to CBPS, residual balancing is computationally
attractive in that the weighting solution is quickly obtained even
with a large number of confounders, time periods, and observations.
An open source R package, \texttt{rbw}, is available for implementing
the proposed method.

In the sections that follow, we first briefly review MSMs and the
method of IPW. Next, we introduce the method of residual balancing,
and conduct a set of simulation studies to evaluate its performance
relative to IPW and its variants. We then illustrate the method empirically
by estimating the cumulative effect of negative advertising on election
outcomes as well as the controlled direct effect (CDE) of shared democracy
on public support for war. We conclude by discussing the method's
limitations along with possible remedies.

\section{MSMs and IPW: A Review}

In this section, we briefly review MSMs and the method of IPW (\citealt{robins1999marginal,robins2000marginal}).
Consider first a study with $T\geq2$ time points where interest is
in the effect of a time-varying treatment, $D_{t}$ ($1\leq t\leq T$),
on an end-of-study outcome, $Y$. At each time point, there is also
a vector of observed time-varying confounders, $X_{t}$, that may
be affected by prior treatments. Following convention, we use overbars
to denote the treatment history, $\overline{D}_{t}=(D_{1},\ldots D_{t})$,
and confounder history, $\overline{X}_{t}=(X_{1},\ldots X_{t})$,
up to time $t$. Similarly, we denote an individual's complete treatment
and confounder histories through the end of follow-up by $\overline{D}=\overline{D}_{T}$
and $\overline{X}=\overline{X}_{T}$, respectively. Finally, we use
$Y(\overline{d})$ to denote the potential outcome under the particular
treatment history $\overline{d}$.

An MSM is a model for the marginal mean of the potential outcomes,
which can be expressed in general form as follows:
\begin{equation}
\mathbb{E}[Y(\overline{d})]=\mu(\overline{d};\beta),\label{eq:msmgeneral}
\end{equation}
where $\mu(\cdot)$ is some function of treatment history, $\overline{d}$,
and a parameter vector, $\beta$, that captures the marginal effects
of interest. For example, with a large number of time points and a
binary treatment, a common parameterization is
\begin{equation}
\mathbb{E}[Y(\overline{d})]=\beta_{0}+\beta_{1}\textup{cum}(\overline{d}),\label{eq:msmcum}
\end{equation}
where $\textup{cum}(\overline{d})=\sum_{t=1}^{T}d_{t}$ denotes the
total number of time periods on treatment and $\beta_{1}$ captures
the marginal effect of one additional wave on treatment. Of course,
many other parameterizations are possible.

An MSM can be identified from observed data under three key assumptions: 
\begin{enumerate}
\item consistency, which requires that, for any unit, if $\overline{D}=\overline{d}$,
then $Y=Y(\overline{d})$; 
\item sequential ignorability, which requires that treatment at each time
point must not be confounded by unobserved factors conditional on
past treatments and observed confounders, or formally, that $Y(\overline{d})\ci D_{t}|\overline{D}_{t-1},\overline{X}_{t}$
for any treatment sequence $\overline{d}$; and 
\item positivity, which requires that treatment assignment must not be deterministic,
or formally, that $f(D_{t}=d_{t}|\overline{D}_{t-1}=\overline{d}_{t-1},\overline{X}_{t}=\overline{x}_{t})>0$
for any treatment condition $d_{t}$ if $f(\overline{D}_{t-1}=\overline{d}_{t-1},\overline{X}_{t}=\overline{x}_{t})>0$,
where $f(\cdot)$ denotes a probability mass or density function. 
\end{enumerate}
When these assumptions are satisfied, an MSM can be consistently estimated
using the method of IPW.

IPW estimation involves fitting a model for the conditional mean of
the observed outcome given an individual's treatment history using
weights that balance, in expectation, past confounders across treatment
at each time point. The inverse probability weight for individual
$i$ is defined as
\begin{equation}
w_{i}=\prod_{t=1}^{T}\frac{1}{f(D_{t}=d_{i,t}|\overline{D}_{t-1}=\overline{d}_{i,t-1},\overline{X}_{t}=\overline{x}_{i,t})},\label{eq:w}
\end{equation}
where the $\overline{D}_{t-1}=\overline{d}_{i,t-1}$ term can be ignored
when $t=1$. Since the denominator of equation \eqref{eq:w} can be
very small, some units may end up with extremely large weights, leading
to highly variable estimates. To mitigate this problem, \citet{robins2000marginal}
suggest using a so-called ``stabilized'' weight, which is defined
as
\begin{equation}
sw_{i}=\prod_{t=1}^{T}\frac{f(D_{t}=d_{i,t}|\overline{D}_{t-1}=\overline{d}_{i,t-1})}{f(D_{t}=d_{i,t}|\overline{D}_{t-1}=\overline{d}_{i,t-1},\overline{X}_{t}=\overline{x}_{i,t})}.\label{eq:sw}
\end{equation}
Sometimes, the probabilities in both the numerator and denominator
are also made conditional on a set of baseline or time-invariant confounders
$C$:

\begin{equation}
sw_{i}=\prod_{t=1}^{T}\frac{f(D_{t}=d_{i,t}|\overline{D}_{t-1}=\overline{d}_{i,t-1},C=c)}{f(D_{t}=d_{i,t}|\overline{D}_{t-1}=\overline{d}_{i,t-1},\overline{X}_{t}=\overline{x}_{i,t},C=c)}.\label{eq:sw2}
\end{equation}
In such cases, these variables need to be included in the MSM to properly
adjust for confounding, which is unproblematic because they cannot
be affected by treatment.

In practice, both the numerator and the denominator of the stabilized
weight need to be estimated. When treatment is binary, the denominator
is typically estimated using a generalized linear model (GLM), with
the logit or probit link function, for treatment at each time point,
while the numerator is estimated using a constrained version of this
model that omits the time-varying confounders. When treatment is continuous,
models are needed to estimate the conditional densities in both the
numerator and the denominator of the weight. After weights have been
computed, the marginal effects of interest are estimated by fitting
a model for the conditional mean of $Y$ given $\overline{D}_{t}$
with weights equal to $sw_{i}$. When both this model and the models
for treatment assignment are correctly specified, this procedure yields
consistent estimates for all marginal means of the potential outcomes,
$\mathbb{E}[Y(\overline{d})]$, and thus for any marginal effect of
interest, provided that the identification assumptions outlined previously
are satisfied.

As shown in prior studies (e.g., \citealt{lefebvre2008impact,howe2011limitation}),
IPW estimates of marginal effects can be highly sensitive to misspecification
of the models used to construct the weights. To address this limitation,
\citet{imai2014covariate,imai2015robust} developed the method of
CBPS to estimate the denominator in equation \eqref{eq:sw} for binary
treatments. With a logit model for treatment at each time point, this
method augments the score conditions of the likelihood function with
a set of covariate balance conditions. Because the number of balance
conditions may exceed the number of model parameters to be estimated,
the generalized method of moments (GMM) is used to minimize imbalance
in the weighted sample. This method of incorporating balance conditions
into model-based estimation of the weights tends to reduce the bias
that results when the treatment models are misspecified (see \citealt{Tan2017regularized}
for a theoretical discussion on the robustness of calibrated propensity
scores).

MSMs and IPW estimation can also be used to examine causal mediation
(\citealt{vanderweele2015explanation}). Consider now a study with
a point-in-time treatment, $D$, a putative mediator measured at some
point following treatment, $M$, and an end-of-study outcome, $Y$.
Suppose that both treatment and the mediator are confounded by a vector
of observed pre-treatment covariates, denoted by $C$, and that the
mediator is additionally confounded by a vector of observed post-treatment
covariates, denoted by $Z$, which may be affected by the treatment
received earlier. In this setting, the potential outcomes of interest
are denoted by $Y(d,m)$.

As before, an MSM models the marginal mean of the potential outcomes.
If, for example, treatment and the mediator are both binary, a saturated
MSM can be expressed as follows:
\begin{equation}
\mathbb{E}[Y(d,m)]=\alpha_{0}+\alpha_{1}d+\alpha_{2}m+\alpha_{3}dm.\label{eq:msmmed}
\end{equation}
From this model, the controlled direct effect of treatment is given
by $\textup{CDE}(m)=\mathbb{E}[Y(1,m)-Y(0,m)]=\alpha_{1}+\alpha_{3}m$,
which measures the strength of the causal relationship between treatment
and the outcome when the mediator is fixed at a given value, $m$,
for all individuals (\citealt{pearl2001direct,robins2003semantics}).
This estimand is useful for assessing causal mediation because it
helps to adjudicate between alternative explanations for a treatment
effect. For example, the difference between a total effect and the
$\textup{CDE}(m)$ may be interpreted as the degree to which the mediator
contributes to a causal mechanism that transmits the effect of treatment
on the outcome (\citealt{acharya2016explaining,zhou-wodtke2018cde}).

MSMs for the joint effects of a treatment and mediator, like equation
\eqref{eq:msmmed}, can be identified under essentially the same assumptions
as outlined previously. In this context, the consistency assumption
requires that $Y=Y(d,m)$ if $D=d$ and $M=m$; sequential ignorability
requires that both treatment and the mediator must be unconfounded
conditional on the observed past, or formally, that $Y(d,m)\ci D|C$
and $Y(d,m)\ci M|C,D,Z$; and positivity requires that both treatment
and the mediator are not deterministic functions of past variables.
Similarly, the stabilized inverse probability weights are here defined
as
\begin{equation}
sw_{i}^{*}=\frac{f(D=d_{i})}{f(D=d_{i}|C=c_{i})}\times\frac{f(M=m_{i}|D=d_{i})}{f(M=m_{i}|C=c_{i},D=d_{i},Z=z_{i})},\label{eq:swmed}
\end{equation}
and they must be estimated using appropriate models for the conditional
probabilities and/or densities that compose this expression. After
weights have been computed, the marginal effects of interest \textendash{}
here, the $\textup{CDE}(m)$ \textendash{} are estimated by fitting
a model for the conditional mean of $Y$ given $D$ and $M$ with
weights equal to $sw_{i}^{*}$. Alternatively, it is also possible
to define the weights as $sw_{i}^{\dagger}=\frac{f(M=m_{i}|C=c_{i},D=d_{i})}{f(M=m_{i}|C=c_{i},D=d_{i},Z=z_{i})}$,
in which case $C$ must be included in the MSM to properly adjust
for confounding. Adjusting for $C$ in the MSM is unproblematic because
these variables are not post-treatment confounders, unlike $Z$.

\section{Residual Balancing}

In this section, we motivate and explain the method of residual balancing.
We first focus on analyses of time-varying treatment effects, and
then we outline how the method is easily adapted for studies of causal
mediation. Finally, we discuss similarities and differences between
residual balancing and the CBPS method proposed by \citet{imai2015robust}.

\subsection{Rationale}

To explain the method of residual balancing, it is useful to begin
with Robins' \citeyearpar{robins1986new} g-computation formula. The
g-computation formula factorizes the marginal mean of the potential
outcome, $Y(\overline{d})$, as follows: 
\begin{equation}
\mathbb{E}[Y(\overline{d})]=\int\cdots\int\mathbb{E}[Y|\overline{D}=\overline{d},\overline{X}=\overline{x}]\prod_{t=1}^{T}f(x_{t}|\overline{x}_{t-1},\overline{d}_{t-1})d\mu(x_{t}).\label{eq:g-computation}
\end{equation}
In contrast, the conditional mean of the observed outcome $Y$ given
$\overline{D}=\overline{d}$ can be factorized into 
\begin{equation}
\mathbb{E}[Y|\overline{D}=\overline{d}]=\int\cdots\int\mathbb{E}[Y|\overline{D}=\overline{d},\overline{X}=\overline{x}]\prod_{t=1}^{T}f(x_{t}|\overline{x}_{t-1},\overline{d})d\mu(x_{t}).\label{eq:observed}
\end{equation}
A comparison of equation \eqref{eq:g-computation} with equation \eqref{eq:observed}
indicates that weighting the observed population by 
\begin{equation}
W_{x}=\prod_{t=1}^{T}\frac{f(X_{t}|\overline{X}_{t-1},\overline{D}_{t-1})}{f(X_{t}|\overline{X}_{t-1},\overline{D})}\label{eq:w2}
\end{equation}
would yield a pseudo-population in which $f^{*}(x_{t}|\overline{x}_{t-1},\overline{d})=f^{*}(x_{t}|\overline{x}_{t-1},\overline{d}_{t-1})=f(x_{t}|\overline{x}_{t-1},\overline{d}_{t-1})$
and thus $\mathbb{E}^{*}[Y|\overline{D}=\overline{d}]=\mathbb{E}^{*}[Y(\overline{d})]=\mathbb{E}[Y(\overline{d})]$,
where the asterisk denotes quantities in the weighted pseudo-population.\footnote{In fact, the ``stabilized'' weight in equation \eqref{eq:sw} is
just a different way of writing equation (10):
\begin{align*}
W_{x} & =\prod_{t=1}^{T}\frac{f(X_{t}|\overline{X}_{t-1},\overline{D}_{t-1})}{f(X_{t}|\overline{X}_{t-1},\overline{D})}=\frac{\prod_{t=1}^{T}f(X_{t}|\overline{X}_{t-1},\overline{D}_{t-1})}{f(\overline{X}|\overline{D})}=\frac{f(\overline{D})\prod_{t=1}^{T}f(X_{t}|\overline{X}_{t-1},\overline{D}_{t-1})}{f(\overline{X},\overline{D})}\\
 & =\text{\ensuremath{\frac{f(\overline{D})\prod_{t=1}^{T}f(X_{t}|\overline{X}_{t-1},\overline{D}_{t-1})}{\prod_{t=1}^{T}f(X_{t}|\overline{X}_{t-1},\overline{D}_{t-1})f(D_{t}|\overline{X}_{t},\overline{D}_{t-1})}}=\ensuremath{\frac{\prod_{t=1}^{T}f(D_{t}|\overline{D}_{t-1})}{\prod_{t=1}^{T}f(D_{t}|\overline{X}_{t},\overline{D}_{t-1})}}}
\end{align*}
} Because $X_{t}$ is often high-dimensional, estimation of the conditional
densities in equation \eqref{eq:w2} is practically difficult.

Nevertheless, the condition that $f^{*}(x_{t}|\overline{x}_{t-1},\overline{d})=f^{*}(x_{t}|\overline{x}_{t-1},\overline{d}_{t-1})=f(x_{t}|\overline{x}_{t-1},\overline{d}_{t-1})$
implies that, in the pseudo-population, the following moment condition
would hold for any scalar function $g(\cdot)$ of $X_{t}$: 
\begin{equation}
\mathbb{E}^{*}[g(X_{t})|\overline{X}_{t-1},\overline{D}]=\mathbb{E}^{*}[g(X_{t})|\overline{X}_{t-1},\overline{D}_{t-1}]=\mathbb{E}[g(X_{t})|\overline{X}_{t-1},\overline{D}_{t-1}].\label{eq:moment1}
\end{equation}
This moment condition can be equivalently expressed as 
\begin{equation}
\mathbb{E}^{*}[\delta(g(X_{t}))|\overline{X}_{t-1},\overline{D}]=0,\label{eq:moment2}
\end{equation}
where $\delta(g(X_{t}))=g(X_{t})-\mathbb{E}[g(X_{t})|\overline{X}_{t-1},\overline{D}_{t-1}]$
is a residual transformation of $g(X_{t})$ with respect to its conditional
mean given the observed past. The moment condition in equation \eqref{eq:moment2}
in turn implies that for any scalar function $h(\cdot)$ of $\overline{X}_{t-1}$
and $\overline{D}$, $\delta(g(X_{t}))$ and $h(\overline{X}_{t-1},\overline{D})$
are uncorrelated, that is, 
\begin{equation}
\mathbb{E}^{*}[\delta(g(X_{t}))h(\overline{X}_{t-1},\overline{D})]=\mathbb{E}^{*}[\delta(g(X_{t}))]\mathbb{E}^{*}[h(\overline{X}_{t-1},\overline{D})]=0,\label{eq:crossmoment}
\end{equation}
where the second equality follows from the fact that $\mathbb{E}^{*}[\delta(g(X_{t}))]=\mathbb{E}^{*}\mathbb{E}^{*}[\delta(g(X_{t}))|\overline{X}_{t-1},\overline{D}]=0$.

The method of residual balancing emulates the moment conditions \eqref{eq:crossmoment}
that would hold in the pseudo-population were it possible to weight
by $W_{x}$. In other words, it emulates the moment conditions \eqref{eq:crossmoment}
that would be expected in a \textit{sequentially} randomized experiment.
Specifically, this is accomplished by (a) specifying a set of $g(\cdot)$
functions, $G(X_{t})=\{g_{1}(X_{t}),\ldots\textit{g}_{\textit{\tiny J}_{t}}(X_{t})\}$,
and a set of $h(\cdot)$ functions, $H(\overline{X}_{t-1},\overline{D})=\{h_{1}(\overline{X}_{t-1},\overline{D}),\ldots h_{\textit{\tiny K}_{t}}(\overline{X}_{t-1},\overline{D})\}$;
(b) computing a set of residual terms, $\delta(g(X_{t}))=g(X_{t})-\mathbb{E}[g(X_{t})|\overline{X}_{t-1},\overline{D}_{t-1}]$,
from the observed data; and then (c) finding a set of weights such
that, for any $j$, $k$, and $t$, the cross-moment of $\delta(g_{j}(x_{it}))$
and $h_{k}(\overline{x}_{i,t-1},\overline{d}_{i})$ is zero in the
weighted data. Hence, it involves finding a set of nonnegative weights,
denoted by $rbw_{i}$, subject to the following balancing conditions:
\begin{equation}
\sum_{i=1}^{n}rbw_{i}\delta(g_{j}(x_{it}))h_{k}(\overline{x}_{i,t-1},\overline{d}_{i})=0,\quad1\leq j\leq J_{t};1\leq k\leq K_{t},\label{eq:constraints}
\end{equation}
or, expressed more succinctly,

\begin{equation}
\sum_{i=1}^{n}rbw_{i}c_{ir}=0,\quad1\leq r\leq n_{c},\label{eq:constraints-1}
\end{equation}
where $c_{ir}$ is the $r$th element of $\bm{c}_{i}=\{\delta(g_{j}(x_{it}))h_{k}(\overline{x}_{i,t-1},\overline{d}_{i});1\leq j\leq J_{t},1\leq k\leq\,K_{t},1\leq t\leq T\}$
and $n_{c}=\sum_{t=1}^{T}J_{t}K_{t}$ is the total number of balancing
conditions. The conditions in equation \eqref{eq:constraints} stipulate
that the residualized confounders at each time point are balanced
across future treatments, past treatments, and past confounders, or
some function thereof. In this way, the proposed method adjusts for
post-treatment confounding without engendering bias due to over-control
or collider-stratification, as the residualized confounders are balanced
across future treatments while (appropriately) remaining orthogonal
to the observed past.

As long as the convex hull of $\{\bm{c}_{i};1\leq i\leq n\}$ contains
$\bm{0}$, finding the weighting solution is an under-identified (or
just-identified) problem. Following \citet{hainmueller2012entropy},
we minimize the relative entropy between $rbw_{i}$ and a set of base
weights $q_{i}$ (e.g., a vector of ones or survey sampling weights),\footnote{Alternative loss functions, such as the empirical likelihood (\citealt{fong2018covariate})
or the variance (\citealt{zubizarreta2015stable}), could also be
used to construct the weights. We use the relatively entropy metric
because it can easily accommodate a set of base weights. Moreover,
in contrast to the empirical likelihood, the relatively entropy metric
is convex and thus computationally convenient. } 
\begin{equation}
\min_{rbw_{i}}\sum_{i}rbw_{i}\log(rbw_{i}/q_{i}),\label{eq:relative entropy}
\end{equation}
subject to the $n_{c}$ balancing conditions. This is a constrained
optimization problem that can be solved using Lagrange multipliers.
Technical details can be found in Supplementary Material A (see also
\citealt{hainmueller2012entropy}).

\subsection{Implementation}

In practice, residual balancing requires specifying a set of $g(\cdot)$
functions that constitute $G(X_{t})$. A natural choice is to set
$g_{j}(X_{t})=X_{jt}$, where $X_{jt}$ is the $j$th element of the
covariate vector $X_{t}$. If there is concern about confounding by
higher-order or interaction terms, they can also be included in $G(X_{t})$.
Then, the residual terms, $\delta(g(X_{t}))$, need to be estimated
from the data. Because $\delta(g(X_{t}))=g(X_{t})-\mathbb{E}[g(X_{t})|\overline{X}_{t-1},\overline{D}_{t-1}]$,
they can be estimated by fitting GLMs for $g(X_{t})$ and then extracting
the response residuals, $\hat{\delta}(g(X_{t}))=g(X_{t})-m(\hat{\beta}_{t}^{T}r(\overline{X}_{t-1},\overline{D}_{t-1}))$,
where $r(\overline{X}_{t-1},\overline{D}_{t-1})=[r_{1}(\overline{X}_{t-1},\overline{D}_{t-1}),\ldots r_{\textit{\tiny L}_{t}}(\overline{X}_{t-1},\overline{D}_{t-1})]$
is a vector of regressors and $m(\cdot)$ denotes the inverse link
function of the GLM.

In addition, residual balancing requires specifying a set of $h(\cdot)$
functions that constitute $H(\overline{X}_{t-1},\overline{D})$. Because
weighting is intended to neutralize the relationship between $X_{t}$
and future treatments, we suggest including all future treatments,
$D_{t}$, $D_{t+1}$,$\ldots$$D_{T}$, in $H(\overline{X}_{t-1},\overline{D})$.
However, if it is reasonable to assume that the effects of $X_{t}$
on future treatments stop at $D_{t'}$, where $t\leq t'<T$, treatments
beyond time $t'$ may be excluded from $H(\overline{X}_{t-1},\overline{D})$.
Equation \eqref{eq:crossmoment} additionally indicates that $\delta(g(X_{t}))$
should be uncorrelated with past treatments, $\overline{D}_{t-1}$,
and past confounders, $\overline{X}_{t-1}$, in the weighted pseudo-population.
Because $\mathbb{E}[\delta(g(X_{t}))|\overline{X}_{t-1},\overline{D}_{t-1}]=0$
by construction, zero correlation is guaranteed in the original unweighted
population, and when the GLMs for $g(X_{t})$ are Gaussian, binomial,
or Poisson regressions with canonical links, the score equations ensure
that the response residuals, $\hat{\delta}(g(X_{t}))$, are orthogonal
to the regressors $r(\overline{X}_{t-1},\overline{D}_{t-1})$ in the
original sample. But to ensure that the response residuals, $\hat{\delta}(g(X_{t}))$,
are also orthogonal to the regressors in the reweighted sample, we
suggest including all members of $r(\overline{X}_{t-1},\overline{D}_{t-1})$
in $H(\overline{X}_{t-1},\overline{D})$.

In general, then, $H(\overline{X}_{t-1},\overline{D})$ should include
all future treatments as well as all regressors in the GLMs for $g(X_{t})$,
including an intercept. A reassuring property of this specification
for $H(\overline{X}_{t-1},\overline{D})$ is that if the GLMs for
$g(X_{t})$ are Gaussian, binomial, or Poisson regressions with canonical
links and they are fit to the weighted sample with all future treatments,
$D_{t},D_{t+1},\ldots D_{T}$, as additional regressors, the coefficients
on future treatments will all be exactly zero and the coefficients
on $r(\overline{X}_{t-1},\overline{D}_{t-1})$ will be the same as
those in the original sample. Therefore, when the GLMs for $g(X_{t})$
are correctly specified, the first moments of $g(X_{t})$ are guaranteed
to be balanced across future treatments, conditional on past treatments
and confounders, as would be expected in a scenario where treatment
is unconfounded by $\overline{X}_{t}$.

In sum, a typical implementation of residual balancing for estimating
the marginal effects of a time-varying treatment proceeds in two steps: 
\begin{enumerate}
\item At each time point $t$ and for each confounder $j$, fit a linear,
logistic, or Poisson regression of $x_{ijt}$, as appropriate given
its level of measurement, on $\overline{x}_{i,t-1}$ and $\overline{d}_{i,t-1}$,
and then compute the response residuals, $\hat{\delta}(x_{ijt})$. 
\item Find a set of weights, $rbw_{i}$, such that: 
\begin{enumerate}
\item in the weighted sample, the residuals, $\hat{\delta}(x_{ijt})$, are
orthogonal to all future treatments and the regressors of $x_{ijt}$;
and 
\item the relative entropy between $rbw_{i}$ and the base weights, $q_{i}$,
is minimized. 
\end{enumerate}
\end{enumerate}
The weighting solution can then be used to fit any MSM of interest.

\subsection{Application to Causal Mediation}

Residual balancing can also be used to estimate an MSM for the joint
effects of a point-in-time treatment, $D$, and mediator, $M$, in
the presence of both pre-treatment confounders, $C$, and a set of
post-treatment confounders, $Z$, for the mediator-outcome relationship.
In this setting, residual balancing is implemented using essentially
the same procedure as outlined previously but with several minor adaptions.
First, for each pre-treatment confounder $j$, compute the response
residuals, $\hat{\delta}(c_{ij})$, by centering it around its sample
mean. Then, for each post-treatment confounder $j$, fit a linear,
logistic, or Poisson regression of $z_{ij}$, depending on its level
of measurement, on $c_{i}$ and $d_{i}$, and then compute the response
residuals, $\hat{\delta}(z_{ij})$. Finally, find a set of weights,
$rbw_{i}$, such that, in the weighted sample, the pre-treatment residuals
$\hat{\delta}(c_{ij})$ are orthogonal to both treatment $d$ and
the mediator $m$, the post-treatment residuals $\hat{\delta}(z_{ij})$
are orthogonal to treatment, the mediator, and the pre-treatment confounders
$c_{ij}$; and the relative entropy between $rbw_{i}$ and the base
weights $q_{i}$ is minimized. The weighting solution can then be
used to fit any MSM for the joint effects of the treatment and mediator
on the outcome, from which the controlled direct effects of interest
are constructed. Alternatively, it is also possible to skip the first
step and construct weights that only balance the residualized post-treatment
confounders, in which case the pre-treatment confounders $C$ must
be included as regressors in the MSM.

\subsection{Comparison with IPW and CBPS}

Compared with IPW, residual balancing has several advantages. First,
because it does not require explicit models for the conditional probability/density
of exposure to treatment and/or a mediator, residual balancing is
robust to the bias that results when these models are misspecified,
and it is easy to use with both binary and continuous exposures. Second,
by minimizing the relative entropy between the balancing weights and
the base weights, the method tends to avoid highly variable and extreme
weights, thus yielding more efficient estimates of causal effects. 

Residual balancing is similar to the CBPS method (\citealt{imai2015robust})
in that it seeks a set of weights that balance time-varying confounders
across future treatments by explicitly specifying a set of balancing
conditions. Residual balancing differs from CBPS, however, in two
important respects. First, unlike CBPS, residual balancing can easily
accommodate continuous treatments and/or mediators. As mentioned previously,
this is because residual balancing does not require parametric models
for exposure to treatment and/or a mediator, and thus it can balance
confounders across both binary and continuous treatments using a common
set of balancing conditions (equation \ref{eq:constraints}). CBPS,
by contrast, is based on a parametric model for the propensity score,
and it is therefore limited to settings with binary treatments and/or
mediators. 

Second, residual balancing allows for the specification of more flexible
and parsimonious balancing conditions than those specified with the
CBPS method. In fact, CBPS can be viewed as an extreme form of residual
balancing. To see the connection, note that CBPS attempts to balance
the time-varying confounders across \textit{all} possible sequences
of future treatments within \textit{each} possible history of past
treatments. Thus, for each confounder $j$, there are $2^{t-1}\times(2^{T-t+1}-1)=2^{T}-2^{t-1}$
balancing conditions at time $t$. Summing over $t$ and $j$, the
total number of balancing conditions associated with CBPS is $n_{c}^{\textup{\tiny CBPS}}=J[(T-1)2^{T}+1]$.
Because $n_{c}^{\textup{\tiny CBPS}}\sim O(J\cdot T\cdot2^{T})$,
it can easily exceed the sample size, in which case the balancing
conditions are at best approximated. With residual balancing, the
number of balancing conditions $n_{c}=\sum_{t=1}^{T}J_{t}K_{t}$ depends
on the choice of $G(X_{t})$ and $H(\overline{X}_{t-1},\overline{D})$.
As mentioned previously, a natural choice of $G(X_{t})$ is $\{X_{1t},X_{2t},\ldots,X_{jt}\}$.
If $\mathbb{E}[g_{j}(X_{t})|\overline{X}_{t-1},\overline{D}_{t-1}]$
is then modeled with a saturated GLM of $X_{jt}$ on $\overline{D}_{t-1}$
only, and $H(\overline{X}_{t-1},\overline{D})$ is defined as a set
of dummy variables for each possible sequence of future treatments
interacted with each possible history of past treatments, the balancing
conditions in equation \eqref{eq:constraints} would be equivalent
to those for the CBPS method. 

With residual balancing, however, $G(X_{t})$, $\mathbb{E}[g_{j}(X_{t})|\overline{X}_{t-1},\overline{D}_{t-1}]$,
and $H(\overline{X}_{t-1},\overline{D})$ can be specified more flexibly.
For example, when a parsimonious GLM is used for $\mathbb{E}[g_{j}(X_{t})|\overline{X}_{t-1},\overline{D}_{t-1}]$,
and only the $L_{t}$ regressors of $g_{j}(X_{t})$ and $T-t+1$ future
treatments are included in $H(\overline{X}_{t-1},\overline{D})$,
the number of balancing conditions will be $n_{c}=J\sum_{t=1}^{T}(T-t+1+L_{t})$,
which is substantially smaller than $n_{c}^{\textup{\tiny CBPS}}$.
In large and even moderately size samples, these balancing conditions
can often be satisfied exactly.

\section{Simulation Experiments}

In this section, we conduct a set of simulation studies to assess
the performance of residual balancing for estimating marginal effects
with (a) a binary time-varying treatment under correct model specification,
(b) a continuous time-varying treatment under correct model specification,
(c) a binary time-varying treatment under incorrect model specification,
and (d) a continuous time-varying treatment under incorrect model
specification. In each of these four settings, we compare residual
balancing with four variants of IPW: conventional IPW with weights
estimated from GLMs (IPW-GLM), IPW with weights estimated from GLMs
and then censored (IPW-GLM-Censored), IPW with weights estimated from
CBPS (IPW-CBPS), and as a benchmark, IPW with weights based on the
true exposure probabilities (IPW-Truth). Because the CBPS method has
not been extended for continuous treatments in the time-varying setting,
we assess the performance of IPW-CBPS only for binary treatments.

The data generating process (DGP) in our simulations is very similar
to that of \citet{imai2015robust}. It involves four time-varying
covariates measured at $T=3$ time periods with a sample of $n=1,000$.
At each time $t$, the covariates $X_{t}$ are determined by treatment
at time $t-1$ and a multiplicative error: $X_{t}=(U_{t}\epsilon_{1t},U_{t}\epsilon_{2t},|U_{t}\epsilon_{3t}|,|U_{t}\epsilon_{4t}|)$,
where $U_{1}=1$, $U_{t}=(5/3)+(2/3)D_{t-1}$ for $t>1$ and $\epsilon_{jt}\sim N(0,1)$
for $1\leq j\leq4$. Treatment at each time $t$ depends on prior
treatment at time $t-1$ and the covariates $X_{t}$. Specifically,
when treatment is binary, it is generated as a Bernoulli draw with
probability $p=\textup{logit}^{-1}[-D_{t-1}+\gamma^{T}X_{t}+(-0.5)^{t}]$,
and when treatment is continuous, it is generated as $D_{t}\sim N(\mu_{t}=-D_{t-1}+\gamma^{T}X_{t}+(-0.5)^{t},\sigma_{t}^{2}=2^{2})$,
where $D_{0}=0$ and $\gamma=\alpha(1,-0.5,0.25,0.1)^{T}$. Here,
we use the $\alpha$ parameter to control the level of treatment-outcome
confounding. We consider two values of $\alpha$, 0.4 and 0.8, corresponding
to scenarios where treatment-outcome confounding is mild and strong,
respectively. Finally, the outcome is generated as $Y\sim N(\mu=250-10\sum_{t=1}^{3}D_{t}+\sum_{t=1}^{3}\delta^{T}X_{t},\sigma^{2}=5^{2})$,
where $\delta=(27.4,13.7,13.7,13.7)^{T}$. To assess the impact of
model misspecification, we use the same DGP, but we recode the ``observed''
covariates as nonlinear transformations of the ``true'' covariates:
specifically, $X_{t}^{*}=(X_{1t}^{3},6\cdot X_{2t},\log(X_{3t}+1),1/(X_{4t}+1))^{T}$.
We then use only the transformed covariates, $X_{t}^{*}$, to implement
IPW, its variants, and residual balancing. Note that the conditional
mean model for $X_{jt}^{*}$ is still correct when the treatment is
binary but incorrect when the treatment is continuous.

For each scenario described previously, we generate 2,500 random samples.
Then, for each sample, we construct weights using IPW-GLM, IPW-GLM-Censored,
IPW-CBPS, and residual balancing. With IPW-GLM, we estimate the weights
using logistic regression for binary treatments and normal linear
models for continuous treatments, assuming homoskedastic errors. With
IPW-GLM-Censored, we follow \citeauthor{cole2008constructing}'s (2008)
example and censor weights at the 1st and 99th percentiles. With IPW-CBPS,
we estimate weights using the methods proposed by \citet{imai2015robust}
with the function \texttt{CBMSM()} in the R package \texttt{CBPS}.
With residual balancing, $G(X_{t})=X_{t}$, and the residual terms
are estimated from linear models for $X_{t}$ with prior treatment
$D_{t-1}$ as a regressor, and $H(\overline{X}_{t-1},\overline{D})$
includes $D_{t}$ as well as the regressors in the model for $X_{t}$
(i.e., 1 and $D_{t-1}$). Finally, with each set of weights, we fit
an MSM by regressing the outcome $Y$ on the three treatment variables
$\{D_{1},D_{2},D_{3}\}$ and denote their coefficient estimates as
$\hat{\beta}_{1}$, $\hat{\beta}_{2}$, and $\hat{\beta}_{3}$. We
obtain the true values of these coefficients by simulating potential
outcomes with the g-computation formula, regressing them on the treatment
variables, and averaging their coefficients over a large number of
simulations. The performance of each method is evaluated using the
simulated sampling distributions of $\hat{\beta}_{1}$, $\hat{\beta}_{2}$,
and $\hat{\beta}_{3}$. 
\begin{figure}[!t]
\centering{}\includegraphics[width=0.75\paperwidth]{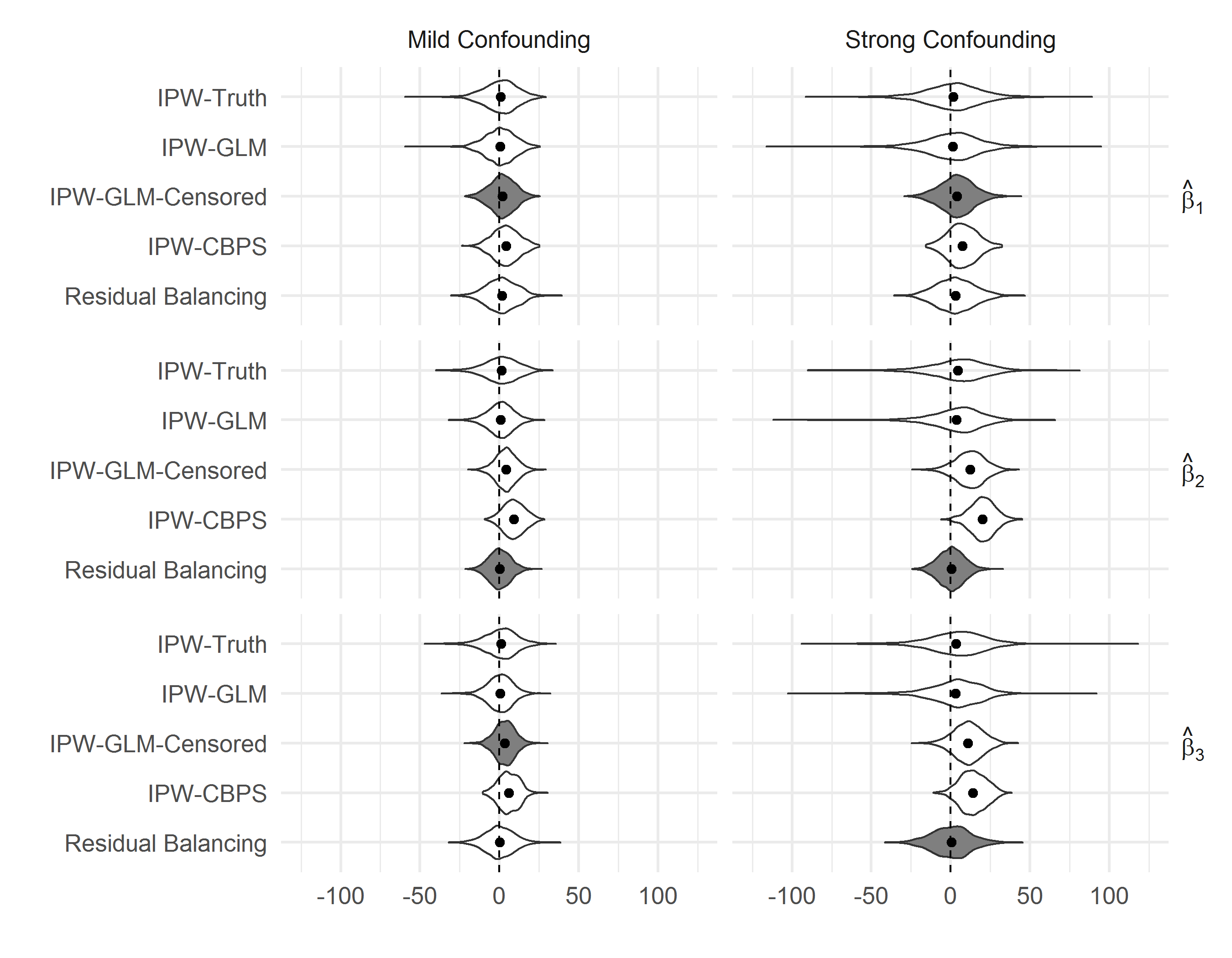}\caption{Simulation results for a binary treatment with correct model specification.
The left and right panels correspond to the settings of ``mild confounding''
($\alpha=0.4$) and ``strong confounding'' ($\alpha=0.8$) respectively.
Four different methods are compared: IPW based on the standard logistic
regression (IPW-GLM), IPW based on the standard logistic regression
with weights censored at the 1st and 99th percentiles (IPW-GLM-Censored),
IPW based on the CBPS (IPW-CBPS), and residual balancing. As a benchmark,
results from IPW based on true treatment probabilities (IPW-Truth)
are also reported. The violin plots show the sampling distributions
(from 2500 random samples) of different estimators centered at the
true values of corresponding parameters, and the shaded violin plots
highlight the estimator with the smallest root mean squared error
(RMSE) in each scenario.}
\end{figure}
\begin{figure}[!t]
\centering{}\includegraphics[width=0.75\paperwidth]{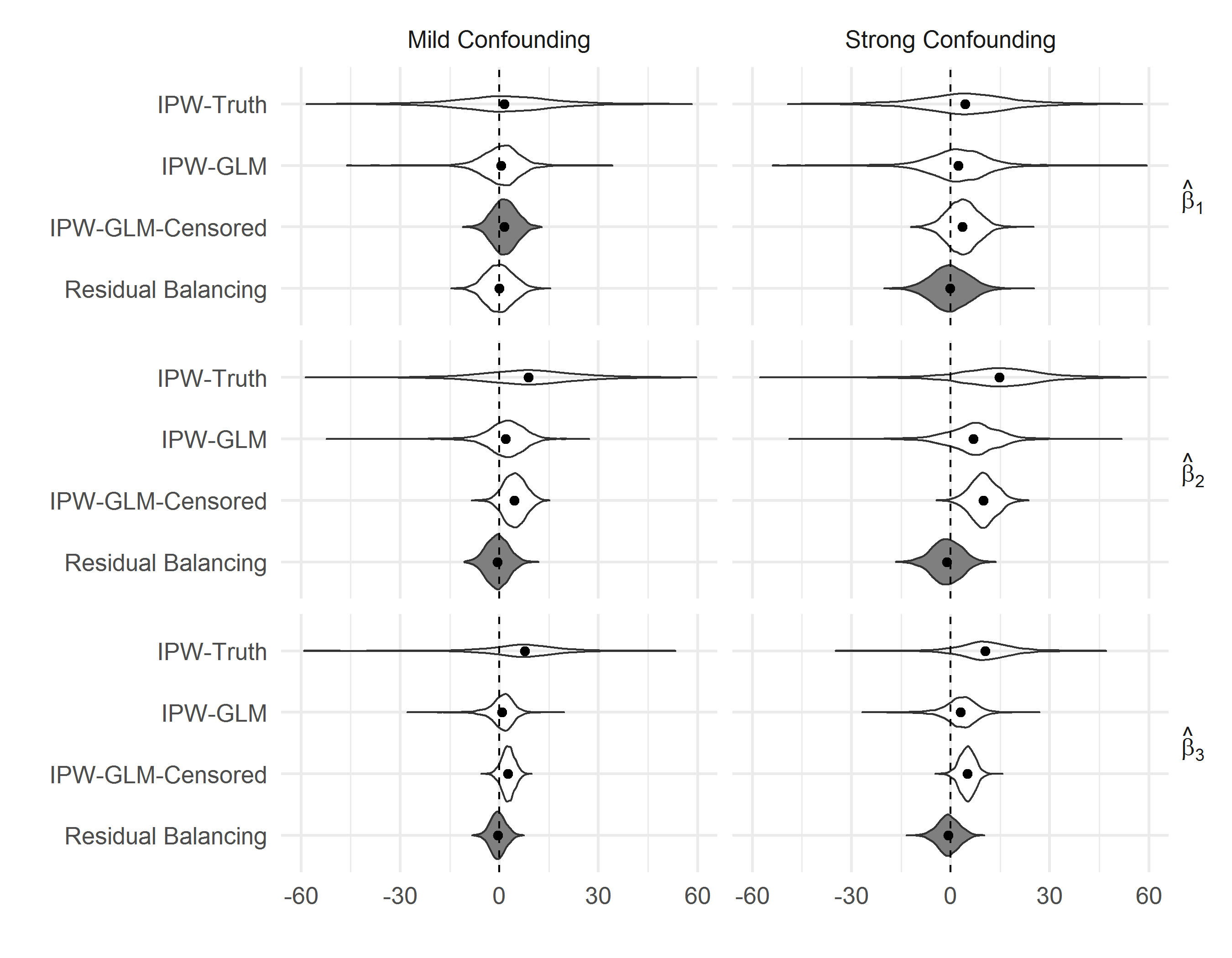}\caption{Simulation results for a continuous treatment with correct model specification.
The left and right panels correspond to the settings of ``mild confounding''
($\alpha=0.4$) and ``strong confounding'' ($\alpha=0.8$) respectively.
Three different methods are compared: IPW based on the standard logistic
regression (IPW-GLM), IPW based on the standard logistic regression
with weights censored at the 1st and 99th percentiles (IPW-GLM-Censored),
and residual balancing. As a benchmark, results from IPW based on
true treatment probabilities (IPW-Truth) are also reported. The violin
plots show the sampling distributions (from 2500 random samples) of
different estimators centered at the true values of corresponding
parameters, and the shaded violin plots highlight the estimator with
the smallest root mean squared error (RMSE) in each scenario.}
\end{figure}
\begin{figure}[!t]
\centering{}\includegraphics[width=0.75\paperwidth]{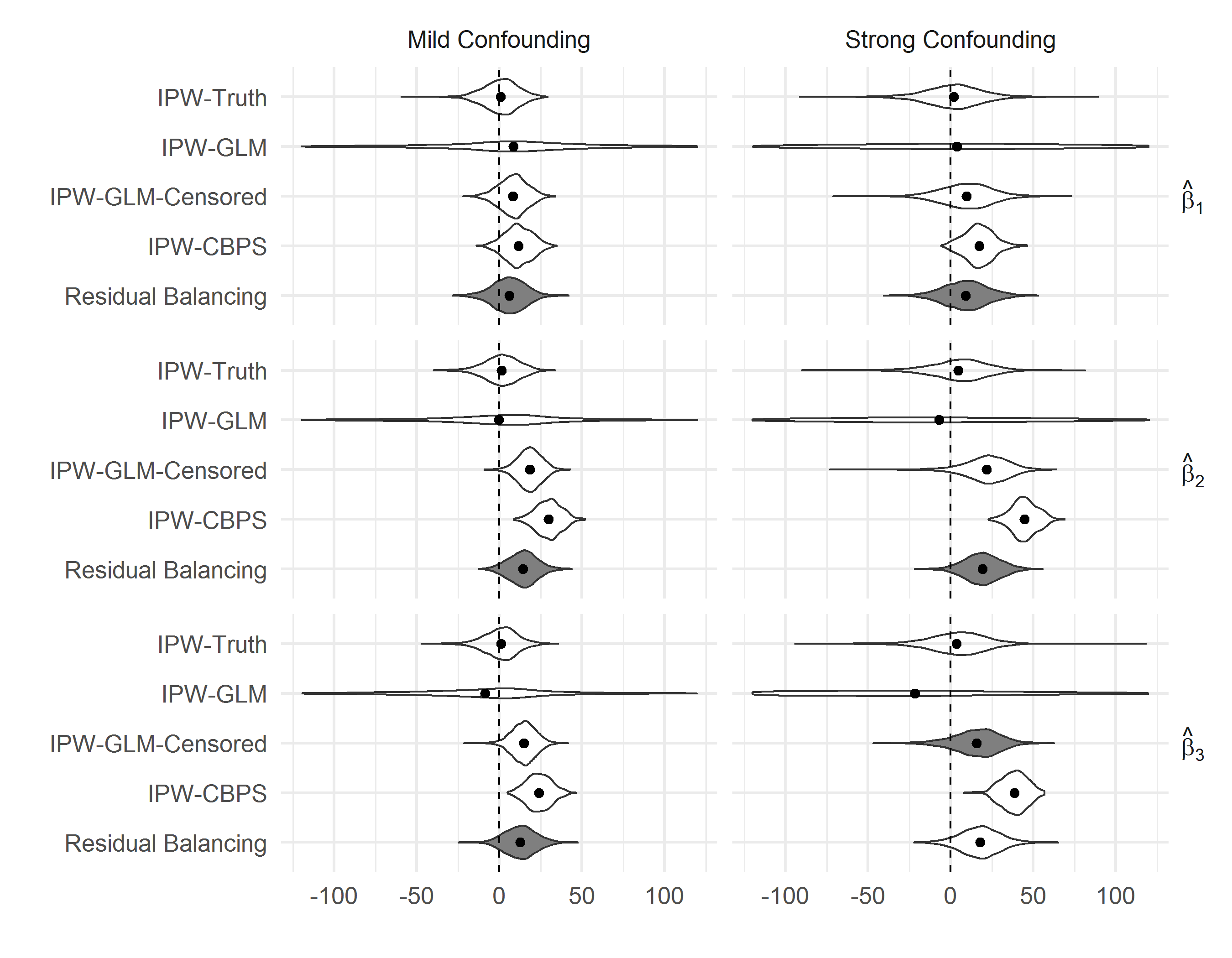}\caption{Simulation results for a binary treatment with incorrect model specification.
The left and right panels correspond to the settings of ``mild confounding''
($\alpha=0.4$) and ``strong confounding'' ($\alpha=0.8$) respectively.
Four different methods are compared: IPW based on the standard logistic
regression (IPW-GLM), IPW based on the standard logistic regression
with weights censored at the 1st and 99th percentiles (IPW-GLM-Censored),
IPW based on the CBPS (IPW-CBPS), and residual balancing. As a benchmark,
results from IPW based on true treatment probabilities (IPW-Truth)
are also reported. The violin plots show the sampling distributions
(from 2500 random samples) of different estimators centered at the
true values of corresponding parameters, and the shaded violin plots
highlight the estimator with the smallest root mean squared error
(RMSE) in each scenario.}
\end{figure}
\begin{figure}[!t]
\centering{}\includegraphics[width=0.75\paperwidth]{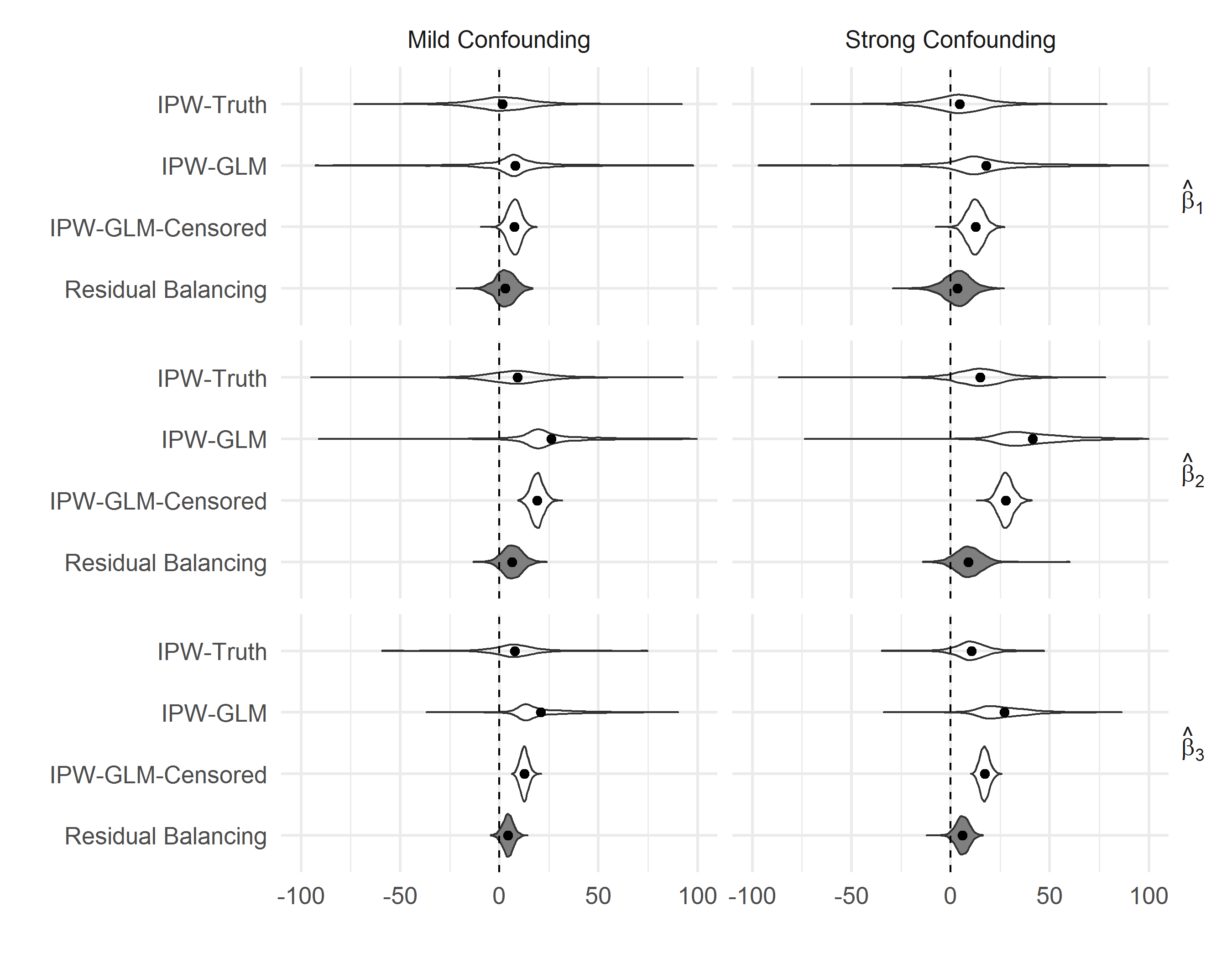}\caption{Simulation results for a continuous treatment with incorrect model
specification. The left and right panels correspond to the settings
of ``mild confounding'' ($\alpha=0.4$) and ``strong confounding''
($\alpha=0.8$) respectively. Three different methods are compared:
IPW based on the standard logistic regression (IPW-GLM), IPW based
on the standard logistic regression with weights censored at the 1st
and 99th percentiles (IPW-GLM-Censored), and residual balancing. As
a benchmark, results from IPW based on true treatment probabilities
(IPW-Truth) are also reported. The violin plots show the sampling
distributions (from 2500 random samples) of different estimators centered
at the true values of corresponding parameters, and the shaded violin
plots highlight the estimator with the smallest root mean squared
error (RMSE) in each scenario.}
\end{figure}

Figure 1 presents results from simulations with a binary treatment
and correct model specification. Specifically, this figure displays
a set of violin plots, which show the sampling distributions of $\hat{\beta}_{1}$,
$\hat{\beta}_{2}$, and $\hat{\beta}_{3}$ centered at the true values
of these coefficients. In these plots, black dots represent means
of the sampling distributions, and the shaded distributions highlight
the estimator with the smallest root mean squared error (RMSE) in
each scenario. 

Two patterns are evident from Figure 1. First, comparing the left
and right panels, we see that IPW and its variants suffer from finite-sample
bias and may have skewed sampling distributions, especially when the
covariates are strongly predictive of treatment. By contrast, residual
balancing is roughly unbiased, and its estimates appear approximately
normally distributed, regardless of the level of confounding. Second,
the results in Figure 1 indicate that residual balancing is much more
efficient than IPW-GLM, especially when the level of confounding is
high. In addition, with a high level of confounding, both IPW-GLM-Censored
and IPW-CBPS yield much less variable estimates than IPW-GLM, but
this gain in precision comes at the expense of greater bias. Residual
balancing, by contrast, improves efficiency without inducing bias.

Figure 2 presents another set of violin plots based on simulations
with a continuous treatment and correct model specification. As before,
the bias for IPW and its variants increases substantially with the
level of confounding. Residual balancing, by contrast, is approximately
unbiased across all levels of confounding. Moreover, residual balancing
consistently outperforms IPW and its variants in terms of efficiency.\footnote{Note that IPW based on true probability densities (IPW-Truth) is often
not as efficient as IPW based on estimated densities (IPW-GLM), which
suggests that ``over-fitting'' the treatment models can lead to
efficiency gains when these models are correctly specified (\citealt{rubin1996matching,hirano2003efficient}).} For example, residual balancing is the most accurate and precise
estimator for $\beta_{2}$ and $\beta_{3}$ under both high and low
levels of confounding (For $\beta_{1}$, the performance of residual
balancing is comparable to that of IPW-GLM-Censored). In sum, residual
balancing matches or exceeds the performance of IPW and its variants
across all scenarios in these simulations.

Figure 3 presents violin plots from simulations with a binary treatment
and misspecified models where $X_{t}$ is measured incorrectly. As
indicated by its extreme level of sampling variation, IPW-GLM is highly
unstable when models for the conditional probability of treatment
are misspecified. Consistent with \citet{imai2015robust}, IPW-CBPS
appears more robust to model misspecification, as reflected in its
substantially smaller sampling variation compared with IPW-GLM. At
the same time, however, this improvement in precision comes at the
cost of greater bias. In addition, censoring the inverse probability
weights also appears to substantially improve the method's performance
in the presence of misspecification. In fact, IPW-GLM-Censored even
outperforms IPW-CBPS in these simulations. Nevertheless, despite the
improvements achieved by censoring the weights or using CBPS, residual
balancing consistently produces the most accurate and efficient estimates
across nearly all scenarios.

Figure 4 presents violin plots from simulations with a continuous
treatment and incorrect measures of $X_{t}$, in which case both the
treatment assignment model for IPW and the confounder models for residual
balancing are misspecified. Consistent with the results discussed
previously, this figure also indicates that IPW-GLM is extremely biased
and inefficient, that censoring the weights reduces bias and improves
efficiency, and that residual balancing yields by far the most accurate
and efficient estimator among all methods. Note that residual balancing
even outperforms IPW based on the true propensity scores, despite
the fact that the confounder models are now misspecified.

\section{The Cumulative Effect of Negative Advertising on Vote Shares}

In this section, we illustrate residual balancing empirically by estimating
the cumulative effect of negative campaign advertising on election
outcomes (\citealt{lau2007effects,blackwell2013framework,imai2015robust}).
Drawing on U.S. senate and gubernatorial elections from 2000 to 2006,
\citet{blackwell2013framework} used MSMs with IPW to evaluate the
cumulative effects of negative campaign advertising on election outcomes
for 114 Democratic candidates. MSMs are appropriate for this problem
because campaign advertising is a dynamic process plagued by post-treatment
confounding. For example, candidates adjust their campaign strategies
on the basis of current polling results, where trailing candidates
are more likely to ``go negative'' than leading candidates. At the
same time, polling results change over time and are likely affected
by a candidate's previous use of negative advertising.

Treatment, $D_{t}$, in this analysis is the proportion of campaign
advertisements that are ``negative'' (i.e., that mention the opposing
candidate) in each campaign-week. Because IPW tends to preform poorly
with continuous treatments, we also consider a binary version of treatment,
\textbf{$B_{t}$}, for which the proportion of negative advertisements
is dichotomized using a cutoff of 10\%, as in \citet{blackwell2013framework}.
The time-varying confounders, $X_{t}$, included in this analysis
are the Democratic share in the polls and the share of undecided voters
in the previous campaign-week. This analysis also uses a set of baseline
confounders, $C$, including total campaign length, election year,
incumbency status, and whether the election is for the senate or governor's
office. The outcome, $Y$, is the Democratic share of the two-party
vote.

Following \citet{imai2015robust}, we focus on the final five weeks
preceding the election and estimate an MSM for the binary version
of treatment with form
\begin{equation}
\mathbb{E}[Y(\overline{b})|C]=\theta_{0}+\theta_{1}\textup{cum}(\overline{b})+\theta_{2}V\cdot\textup{cum}(\overline{b})+\theta_{3}^{T}C,\label{eq:binary}
\end{equation}
and an MSM for the continuous treatment with form
\begin{equation}
\mathbb{E}[Y(\overline{d})|C]=\beta_{0}+\beta_{1}\textup{ave}(\overline{d})+\beta_{2}V\cdot\textup{ave}(\overline{d})+\theta_{3}^{T}C.\label{eq:continuous}
\end{equation}
In these models, $\textup{cum}(\overline{b})$ denotes the total number
of campaign-weeks for which more than 10\% of the candidate's advertising
was negative, $\textup{ave}(\overline{d})$ denotes the average proportion
of advertisements that were negative over the final five weeks of
the campaign, $V$ is an indicator of incumbency status used to construct
interaction terms that allow the effect of negative advertising to
differ between incumbents and nonincumbents.\footnote{In equations \eqref{eq:binary} and \eqref{eq:continuous}, the ``main''
effect of $V$ is captured in the term $\theta_{3}^{T}C$.} Thus, the effect of an additional week with more than 10\% negative
advertising for nonincumbents is $\theta_{1}$, and for incumbents,
it is $\theta_{1}+\theta_{2}$. Similarly, $\beta_{1}$ and $\beta_{1}+\beta_{2}$
correspond to the effects of a 1 percentage point increase in negative
advertising for nonincumbents and incumbents, respectively. To facilitate
comparison of results across the different versions of treatment,
we report estimates for the effects of a 10 percentage point increase
in negative advertising\textemdash that is, $10\beta_{1}$ and $10(\beta_{1}+\beta_{2})$.

We estimate these models with both IPW methods and residual balancing.
Specifically, we first implement IPW-GLM by fitting, at each time
point, a logistic regression of the dichotomized treatment on both
time-varying confounders and baseline confounders, and then constructing
the inverse probability weights using equation \eqref{eq:sw2}. Second,
we implement IPW-CBPS with the same treatment assignment model using
the function \texttt{CBMSM()} in the R package \texttt{CBPS}. Finally,
we implement residual balancing by, first, fitting linear models for
each covariate in $X_{t}$ ($t\geq2$) with lagged values of treatment
and the time-varying confounders as regressors and extracting residual
terms $\hat{\delta}(X_{t})$. For each covariate in $X_{1}$, the
residual term is computed as the deviation from its sample mean. Then,
we find a set of minimum entropy weights such that, in the weighted
sample, $\hat{\delta}(X_{t})$ is balanced across treatment at time
$t$ and the regressors of $X_{jt}$. Standard errors are computed
using the robust (i.e., ``sandwich'') variance estimator.\footnote{In Part B of the Supplementary Material, we report a set of simulation
results on the performance of the robust variance estimator for IPW-GLM,
IPW-GLM-Censored, IPW-CBPS, and residual balancing. We find that the
robust variance estimator is consistently conservative for residual
balancing. For IPW and its variants, the robust variance estimator
appears to sometimes over-estimate and other times under-estimate
the true sampling variance, depending on the particular scenario.} R code for implementing residual balancing in this analysis is available
in Part C of the Supplementary Material.

\begin{table}[!]
\caption{Estimated Marginal Effects of Negative Advertising on the Candidate's
Vote Share}
{\footnotesize{}\smallskip{}
}{\footnotesize\par}
\noindent \begin{centering}
\begin{tabular}{ccccc}
\hline 
\noalign{\vskip0.1cm}
\multirow{2}{*}{Estimator} & \multicolumn{2}{c}{Dichotomized Treatment} & \multicolumn{2}{c}{Continuous Treatment}\tabularnewline[0.1cm]
\noalign{\vskip0.1cm}
 & Nonincumbent & Incumbent & Nonincumbent & Incumbent\tabularnewline[0.1cm]
\hline 
\noalign{\vskip0.1cm}
IPW-GLM & 1.42 (0.43) & -1.73 (0.47) & 0.80 (0.28) & -1.15 (0.31)\tabularnewline[0.1cm]
\noalign{\vskip0.1cm}
IPW-CBPS & 0.78 (0.89) & -2.03 (0.41) &  & \tabularnewline[0.1cm]
\noalign{\vskip0.1cm}
Residual Balancing & 0.98 (0.54) & -1.67 (0.46) & 0.49 (0.32) & -0.99 (0.36)\tabularnewline[0.1cm]
\hline 
\end{tabular}{\footnotesize{}\smallskip{}
}{\footnotesize\par}
\par\end{centering}
\noindent \hangindent=0.3cm\hangafter=0 Note: For the dichotomized
treatment, results represent the estimated marginal effects of an
additional week with more than 10\% negative advertising. For the
continuous treatment, results represent the estimated marginal effects
of a 10 percentage point increase in the average proportion of negative
advertisements across all campaign-weeks. Numbers in parentheses are
robust (i.e., ``sandwich'') standard errors.
\end{table}
Results from these analyses are presented in Table 1, where the first
two columns contain IPW-GLM, IPW-CBPS, and residual balancing estimates
based on the dichotomized version of treatment. For nonincumbent candidates,
these results suggest that the effect of negative advertising is positive.
However, both IPW-CBPS and residual balancing yield point estimates
that are considerably smaller than IPW-GLM. While IPW-GLM suggests
that an additional week with more than 10\% negative advertising increases
a candidate's vote share by 1.42 percentage points, on average, the
estimated effect is reduced to 0.78 percentage points for IPW-CBPS
and 0.98 percentage points for residual balancing. For incumbent candidates,
all three methods indicate that negative advertising has a substantively
large negative effect on vote shares. Residual balancing, for example,
suggests that an additional week with more than 10\% negative advertising
decreases a candidate's vote share by 1.67 percentage points, on average.

The last two columns of Table 1 present results based on the continuous
version of treatment. Because IPW-CBPS has not been developed for
continuous treatments in the time-varying setting, we focus on estimates
from IPW-GLM and residual balancing. Overall, these results are quite
consistent with those based on the dichotomized treatment. For nonincumbents,
the effect of negative advertising appears to be positive, although
the estimate from residual balancing is relatively small. For incumbents,
both methods suggest a sizable negative effect. According to the residual
balancing estimate, a 10 percentage point increase in the proportion
of negative advertising throughout the final five weeks of the campaign
reduces a candidate's vote share by about one percentage point, on
average.

\section{The Controlled Direct Effect of Shared Democracy on Public Support
for War}

In this section, we reanalyze data from \citet{tomz2013dempeace}
to estimate the controlled direct effect (CDE) of shared democracy
on public support for war, controlling for a respondent's perceived
morality of war. With a nationally representative sample of 1,273
US adults, \citet{tomz2013dempeace} conducted a survey experiment
to analyze the role of public opinion in the democratic peace, that
is, the empirical regularity that democracies almost never fight each
other. In this experiment, they presented respondents with a situation
in which a country was developing nuclear weapons and, when describing
the situation, they randomly and independently varied three characteristics
of the country: political regime (whether it was a democracy), alliance
status (whether it had signed a military alliance with the United
States), and economic ties (whether it had high levels of trade with
the United States). They then asked respondents about their levels
of support for a preventive military strike against the country's
nuclear facilities. The authors found that individuals are substantially
less supportive of military action against democracies than against
otherwise identical autocracies.

To investigate the causal mechanisms through which shared democracy
reduces public support for war, \citet{tomz2013dempeace} also measured
each respondent's beliefs about the threat posed by the potential
adversary (\textit{threat}), the cost of military intervention (\textit{cost}),
and the likelihood of victory (\textit{success}). In addition, the
authors also assessed each respondent's moral concerns about using
military force (\textit{morality}). With these data, they conducted
a causal mediation analysis and found that shared democracy reduces
public support for war primarily by changing perceptions of the threat
and morality of using military force. In this analysis, the authors
examined the role of each mediator separately by assuming that they
operate independently and do not influence one another. However, it
is likely that one's perception of morality is partly influenced by
beliefs about the threat, cost, and likelihood of success, which also
affect support for war directly.\footnote{\citet{tomz2013dempeace} acknowledged this possibility and, in an
auxiliary analysis, they relaxed this assumption and estimated the
indirect effect through morality that is not mediated by other mediators,
i.e., the path-specific effect of democracy$\to$morality$\to$support
for war. However, because this effect excludes other pathways through
morality (such as democracy$\to$perceived threat$\to$morality$\to$support
for war), it does not fully capture the mediating role of morality.} Thus, in the following analysis, we treat these variables as post-treatment
confounders and reassess the mediating role of morality accordingly.

In this data set, the outcome, $Y$, is a measure of support for war
on a five-point scale; treatment, $D$, denotes whether the country
developing nuclear weapons is presented as a democracy; the mediator,
$M$, is a dummy variable indicating whether the respondent thought
it would be morally wrong to strike; the pretreatment covariates $C$
include dummy variables for each of the two other randomized treatments
(alliance status and economic ties) as well as a number of demographic
and attitudinal controls; and the post-treatment confounders $Z$
include measures of the respondent's beliefs about threat, cost, and
likelihood of success.\footnote{For detailed descriptions of the variables included in $X$ and $Z$,
see \citet[Table 5]{tomz2013dempeace}.} We estimate the CDE of shared democracy, controlling for perceptions
of morality, using an MSM with form
\begin{equation}
\mathbb{E}[Y(d,m)|C]=\alpha_{0}+\alpha_{1}d+\alpha_{2}m+\alpha_{3}dm+\alpha_{4}^{T}C.\label{eq:immigr}
\end{equation}
In this model, we control for the pretreatment covariates because,
although treatment is randomly assigned, they may still confound the
mediator-outcome relationship.\footnote{Alternatively, these pretreatment confounders can be adjusted for
using IPW or residual balancing weights. We adjust for them directly
in the MSM for the sake of statistical efficiency.} The controlled direct effect is given by $\textup{CDE}(m)=\alpha_{1}+\alpha_{3}m$,
where $\alpha_{1}$ measures the effect of shared democracy on support
for war if none of the respondents had moral reservations about military
intervention and $\alpha_{1}+\alpha_{3}$ measures the effect of shared
democracy on support for war if all respondents thought it would be
morally wrong to strike.

We estimate this model with both IPW-GLM and residual balancing weights.
Specifically, we first implement IPW-GLM by fitting a logit model
for $M$ with $C$, $D$, and $Z$ as regressors, by fitting a second
logit model for $M$ with only $C$ and $D$ as regressors, and then
by using the fitted values from these models to estimate a set of
weights with the following form: $sw_{i}^{\dagger}=\frac{\mathbb{P}(M=m_{i}|C=c_{i},D=d_{i})}{\mathbb{P}(M=m_{i}|C=c_{i},D=d_{i},Z=z_{i})}$.
Second, we implement residual balancing by fitting a linear model
for each post-treatment confounder in $Z$ with $C$ and $D$ as regressors,
computing residual terms $\hat{\delta}(Z)$, and then finding a set
of minimum entropy weights such that, in the weighted sample, $\hat{\delta}(Z)$
is balanced across $M$ and the regressors of $Z$. Standard errors
are computed using the robust (i.e., ``sandwich'') variance estimator.
R code for implementing residual balancing in this analysis is available
in Part C of the Supplementary Material.

\begin{table}[!]
\caption{Estimated CDE of Shared Democracy on Support for War using IPW and
Residual Balancing}
\smallskip{}

\begin{centering}
\begin{tabular}{l>{\centering}m{2.5cm}>{\centering}m{2.5cm}>{\centering}m{3.5cm}}
\hline 
\noalign{\vskip0.1cm}
 & Total Effect & IPW & Residual Balancing\tabularnewline[0.1cm]
\hline 
\noalign{\vskip0.1cm}
\multirow{1}{*}{intercept} & 2.39 (0.05) & 3.12 (0.05) & 2.76 (0.05)\tabularnewline[0.1cm]
\noalign{\vskip0.1cm}
\multirow{1}{*}{shared democracy } & -0.35 (0.07) & -0.20 (0.07) & -0.36 (0.08)\tabularnewline[0.1cm]
\noalign{\vskip0.1cm}
\multirow{1}{*}{moral concerns} &  & -1.63 (0.14) & -1.20 (0.13)\tabularnewline[0.1cm]
\noalign{\vskip0.1cm}
\multirow{1}{*}{shared democracy {*} moral concerns} &  & -0.05 (0.16) & 0.14 (0.16)\tabularnewline[0.1cm]
\hline 
\end{tabular}
\par\end{centering}
\smallskip{}
Note: Coefficients of pretreatment covariates are omitted. For ease
of interpretation, all pretreatment covariates are centered at their
means. Numbers in parentheses are robust (``sandwich'') standard
errors.
\end{table}
As a benchmark, the first column of Table 2 presents an estimate of
the total treatment effect from a regression of $Y$ on $X$ and $A$.
Echoing the original study, we find that shared democracy significantly
reduces public support for war\textemdash by 0.35 points on the five-point
scale, or about 0.25 standard deviations. The next two columns present
IPW and residual balancing estimates, respectively, for model \eqref{eq:immigr}.
In this model, the ``main effect'' of shared democracy represents
the estimated CDE if respondents had no moral reservations about military
intervention, and the sum of this coefficient and the interaction
term represents the estimated CDE if respondents did have moral reservations. 

IPW and residual balancing yield somewhat different estimates of these
effects. According to IPW, the estimated CDE of shared democracy is
-0.20 if respondents had no moral concerns about war, and it is -0.25
if respondents thought it was morally wrong to strike. According to
residual balancing, by contrast, the estimated CDE of shared democracy
is -0.36 if respondents had no moral concerns about war, and it is
-0.22 if respondents thought military intervention was morally wrong.
Notwithstanding these differences, however, both IPW and residual
balancing suggest that most of the total effect is ``direct,'' ,
transmitted through pathways other than morality.

\section{Discussion and Conclusion}

Post-treatment confounding arises in analyses of both time-varying
treatments and causal mediation, where it complicates the use of conventional
regression and matching methods for causal inference. To adjust for
this type of confounding, researchers most often use MSMs along with
the associated method of IPW estimation (\citealt{robins1999marginal,robins2000marginal,vanderweele2015explanation}).
IPW, however, is highly sensitive to model misspecification, relatively
inefficient, susceptible to finite-sample bias, and difficult to use
with continuous treatments. Several remedies for these problems have
been proposed, such as censoring the weights (\citealt{cole2008constructing})
or constructing them with CBPS (Imai and Ratkovic \citeyear{imai2014covariate,imai2015robust}),
but these corrections are not without their own limitations.

In this article, we proposed the method of residual balancing for
constructing weights that can be used to estimate MSMs. In contrast
to IPW, residual balancing does not require models for the conditional
distribution of exposure to treatment and/or a mediator. Rather, it
entails modeling only the conditional means of the post-treatment
confounders, and because it simultaneously imposes covariate balancing
and minimum entropy conditions on the weights, the method is both
more efficient and more robust to model misspecification than IPW.
It is also much easier to use with continuous treatments, which obviates
the need for arbitrary quantile binning as is often employed in practice
(e.g., \citealt{wodtke2011neighborhood,blackwell2013framework}). 

Residual balancing appears to outperform IPW even when the weights
are constructed with CBPS, which similarly incorporate explicit balancing
conditions when estimating the conditional probabilities of exposure.
The reason, we believe, is that IPW with CBPS attempts to balance
the time-varying confounders across all possible sequences of future
treatments within all possible histories of prior treatments, whereas
residual balancing models the conditional means of the time-varying
confounders and balances only their residuals across a parsimonious
representation of future treatments and the observed past. As a result,
the search for covariate balancing weights is often an over-identified
problem with CBPS but an under-identified problem with residual balancing.
Thus, although weights based on CBPS can improve covariate balance
compared with weights estimated from conventional GLMs, the weights
given by residual balancing can satisfy a set of balancing conditions
exactly. 

Despite its many advantages, residual balancing is still limited in
several ways. First, it requires modeling the conditional means of
the post-treatment confounders (or transformations thereof). When
these models are misspecified, the moment condition in equation \eqref{eq:moment1}
is only partially achieved. In this case, equation \eqref{eq:moment2}
implies 
\[
\mathbb{E}^{*}[g(X_{t})|\overline{X}_{t-1},\overline{D}]=\mathbb{E}^{*}[g(X_{t})|\overline{X}_{t-1},\overline{D}_{t-1}]\neq\mathbb{E}[g(X_{t})|\overline{X}_{t-1},\overline{D}_{t-1}],
\]
where future treatments (i.e., $D_{t}$, $D_{t+1}$,$\ldots$$D_{T}$)
may still be unconfounded in the weighted pseudo-population but the
pseudo-population no longer mimics the original unweighted population.
As a result, estimates of marginal effects based on residual balancing
weights may not be consistent for the target population of interest.
This limitation can be mitigated in practice, however, by fitting
more flexible, non- or semi-parametric models for $\mathbb{E}[g(X_{t})|\overline{X}_{t-1},\overline{D}_{t-1}]$.

Second, even when models for the conditional means of the post-treatment
confounders are correctly specified, residual balancing estimates
of marginal effects may still be biased if the balancing conditions
are insufficient. For example, if both the treatment and outcome are
affected by the product of two confounders, say $X_{1t}X_{2t}$, but
$X_{1t}$ and $X_{2t}$ are only included separately in the implementation
of residual balancing, confounding may still be present in the weighted
sample, leading to bias. This bias, however, can be mitigated by including
a large set of functions in $G(X_{t})$, such as $X_{1t}X_{2t}$ along
with other cross-product or higher-order terms. Alternatively, subject
matter knowledge should guide the choice of functions in $G(X_{t})$
when available.

In sum, residual balancing provides an efficient and robust method
of constructing weights for MSMs. It should therefore find wide application
in analyses of time-varying treatments and causal mediation, wherever
post-treatment confounding presents itself. To facilitate its implementation
in practice, we have developed an open-source R package, \texttt{rbw},
for constructing residual balancing weights, which is available from
GitHub: \href{https://github.com/xiangzhou09/rbw}{https://github.com/xiangzhou09/rbw}.
In addition, Part C of the Supplementary Material provides R code
illustrating the use of \texttt{rbw} in our two empirical examples.
\clearpage\bibliographystyle{apsr}
\bibliography{causality_ref}

\end{document}